\documentclass[final,1p,times]{elsarticle}
\biboptions{sort&compress}
\usepackage{amssymb}
\usepackage{amsthm}
\usepackage{subfig}
\usepackage{amsmath}
\usepackage{floatrow}
\usepackage{url}
\journal{Physica A}
\begin{document}


\begin{frontmatter}

\title{The fundamental Diagram of Pedestrian Model with Slow Reaction}

\author[L1,L2]{Jun Fang\corref{cor1}}
\cortext[cor1]{Corresponding author. E-mail address: fangjun06@mails.tsinghua.edu.cn}
\author[L1,L2]{Zheng Qin}
\author[L1,L2]{Hao Hu}
\author[L1,L2]{Zhaohui Xu}
\author[L3]{Huan Li}
\address[L1]{Department of Computer Science and Technology, Tsinghua University, Beijing 100084, China}
\address[L2]{Tsinghua National Laboratory for Information Science and Technology, Tsinghua University, Beijing 100084 , China}
\address[L3]{Department of Computer Science and Technology, Dongguan University of Technology, Dongguan 523808, China}

\begin{abstract}
The slow-to-start models are a classical cellular automata model in simulating vehicle traffic. However, to our knowledge, the slow-to-start effect has not considered in modeling pedestrian dynamic. We verify the similar behavior between pedestrian and vehicle, and propose an new lattice gas (LG) model called the slow reaction (SR) model to describe the pedestrian's delayed reaction in single-file movement. We simulate and reproduce the Seyfried's field experiments at the research centre J\"{u}lich, and use its empirical data to validate our SR model. We compare the SR model with the standard LG model. We test different probability of slow reaction $p_s$ in SR model and found the simulation data of $p_s=0.3$ fit the empirical data best. The RMS error of mean velocity of SR model is smaller than that of standard LG model. In the range of $p_s=0.1\sim0.3$, our fundamental diagram between  velocity and density by simulation coincides with field experiments.  The distribution of  individual velocity in fundamental diagram in SR model agrees with the empirical data better than that of standard LG model. In addition, we observe the stop-and-go waves and phase separation in pedestrian flow by simulation. We reproduced the phenomena of uneven distribution of interspaces by SR model while the standard LG model did not implement. The SR model can reproduce the evolution of spatio-temporal structures of pedestrian flow with higher fidelity to Seyfried's experiments than the standard LG model.
\end{abstract}


\end{frontmatter}

\section{Introduction}
In recent years, the modeling of pedestrian flows has become one of the most exciting topics. It has attracted considerable attention from the physical science, traffic engineering, computer engineering, or even social psychology science \cite{Schadschneider2009_3142}. The lattice gas (LG) model, as a cellular automata (CA) model, is a simple but effective method to simulate the movement of a large number of pedestrians. With very low cost in terms of computer simulation time, it is able to simulate the empirical results qualitatively, even quantitatively with acceptable error. It can reproduce several typical self-organization phenomena observed in real pedestrian flow steadily, such as phase transition, scaling behavior, faster-is-slower effect, stop-and-go wave and lane formation.

Muramatsu introduced LG model to simulate the counter flow in the channel of subway in earlier times \cite{Muramatsu1999_487}. Jiang extended the standard LG model through setting the maximum velocity more than one cell per step \cite{Jiang2007_683}. Kuang and Fukamachi studied the binary mixture flow with two different velocities in their own paper \cite{Kuang2008_066117, Fukamachi2007_425}. Some models introduced the surrounding environment perception and extended pedestrians' visual field to longer distance \cite{Tajima2002_709, Yu2007_046112, Maniccam2006_512}. Yu and Yang considered the right-moving reference of Chinese pedestrians in their own paper \cite{Yu2007_026102, Yang2008_3281}. Li and Fukamachi introduced the behavior of position exchange and sidle into the LG model \cite{Li2005_619, Fukamachi2007_269}. Weng divided pedestrian movement into three basic behaviors, i.e., move, avoid and swirl \cite{Weng2007_668}. The models extension of Matsui \cite{Matsui2009_157} and Song \cite{Song2006_492} belong to finer discretization or multi-grid and Song's model incorporated the social force model \cite{Helbing2000_487} into LG model. Some modified models dealt with evacuation dynamic in the state of emergency, such as crawl behavior in fire disaster \cite{Nagai2006_449} and indoor evacuation without visibility \cite{Nagatani2004_638}.

Calibrating and validating pedestrian models are essential to flow prediction and pedestrian facilities management. However, so far there have been only limited attempt about these work. Seyfried and his colleagues at research centre J\"{u}lich conducted many solid field experiments since the mid-2000s \cite{Seyfried2005_P10002, Seyfried2006_232, Schadschneider2011_557,  Portz2011_577}, which were performed within the Hermes projects \cite{Website:Seyfried}. The trajectories of all pedestrians were generated automatically from video footage \cite{Boltes2010_43}, which can be used for the calibration.

The slow-to-start models are a classical CA model in simulating motor traffic, which were proposed in the 1990s and researched broadly after then. The related papers have been quoted extensively, such as \cite{Takayasu1993_860, Benjamin1996_3119, Barlovic1998_793}. However, to our knowledge, the slow-to-start effect has never applied to pedestrian dynamic modeling. Although the difference of velocity and size between the vehicle and pedestrian is one or two orders of magnitude, we think some similar behavior exists in pedestrian movement. We call this effect the slow reaction of pedestrian. 
From \cite{Seyfried2005_P10002} and videos in web site~\cite{Website:julich}, we observe that for pedestrian flow at high density, the individual velocity is small and the concentration of pedestrians is sometimes reduced. In this case, pedestrian often make delayed reaction on the movement of the pedestrian in front and cannot  follow the front   pedestrian immediately.  Similar phenomena also occur in highways traffic \cite{Takayasu1993_860, Benjamin1996_3119, Barlovic1998_793} despite within a smaller spatio-temporal structure. In addition, The quickness degree of reaction varies from pedestrian to pedestrian. Not all pedestrians within a group react to movement of front pedestrians simultaneously.  Some individuals leave much larger than average gaps in front and the distribution of individual interspace is uneven. This phenomena is also reported in \cite{Seyfried2010_496} and is reproduced by social force model in \cite{Seyfried2005_357} simulating the same scenario with our work.

Based on above reasons, we propose an new LG model called the slow reaction (SR) model to describe the pedestrian's delayed reaction. We try to reproduce Seyfried's experiment of circular passageway about single-file pedestrian flow \cite{Seyfried2005_P10002} and use its empirical data to validate our SR model. In addition, we compare our simulation data with standard LG model to justify our SR model of pedestrian flow.

\section{Models}

\begin{figure}[!ht]
\centering
\includegraphics[width=0.5\textwidth]{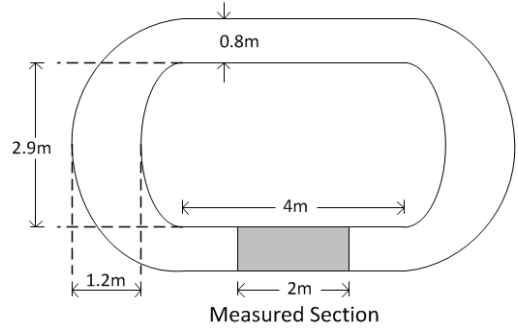}
\caption{Seyfried's experimental set-up in \cite{Seyfried2005_P10002}.}
\label{fig1}
\end{figure}

\begin{figure}[!ht]
\centering
\includegraphics[width=\textwidth]{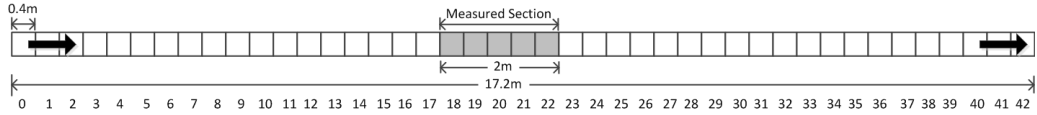}
\caption{Virtual passageway of our simulation with periodic boundary.}
\label{fig2}
\end{figure}

Our model is designed to reproduce the single-file movement in an approximate elliptical circle, which was the experimental set-up in Seyfried's field experiments in \cite{Seyfried2005_P10002} (see Fig.~\ref{fig1}). The experimental area is transformed to a rectangular grid with periodic boundary as our virtual passageway (see Fig.~\ref{fig2}). The virtual passageway is composed of 43 square cells with size of 0.4m*0.4m, corresponding to total length $l_p=17.2$m in physical unit (a litter shorter than 17.3m in \cite{Seyfried2005_P10002}). The cell size of our model is the same as the standard LG model \cite{Schadschneider2009_3142} and can hold one virtual pedestrian at each time step.  At the center of passageway, there is a measured section with length $l_m=5$cell, corresponding to the $ l_m=2$m in physical unit, just like in \cite{Seyfried2005_P10002}. All pedestrians move in single-file pattern and are forbidden to overpass. Therefore, the width of virtual passageway is insignificant in simulation.  As doing in \cite{Seyfried2005_P10002}, we calculate the one-dimensional density $\rho_{1d}$ in the longitudinal distance from the measured section.

\renewcommand{\figurename}{Fig.}
\captionsetup[figure]{labelsep=period}

\subsection{Standard LG model}

In LG model, each one moves one cell at most per time step. The positions of all pedestrians are updated in parallel style.  The states of pedestrians in standard LG model can be divided into two classes according to the relative positions of successive pedestrians: the free state and the block state. For Free state, the gap between the target pedestrian and its front pedestrian is not less than one cell. We adopt the free moving velocity $v_{\rm{free}}=1.24$m/s in \cite{Seyfried2005_P10002} as our calibration parameter. Therefore, one simulation time step corresponds to 0.4/1.24=0.32s in physical unit.  The pedestrian gets into the block state when there is no gap between it and its front one.

\subsection{SR model}
For pedestrian flow at high density, the individual velocity is small and the concentration of pedestrians is sometimes reduced. In this case, pedestrian often make delayed reaction on the movement of the pedestrian in front and cannot  follow the front  pedestrian immediately. In addition, the quickness degree of reaction varies from pedestrian to pedestrian. Not all pedestrians within a group react to movement of front pedestrians simultaneously. Some individuals leave much larger than average gaps in front at the border of and the distribution of individual interspace is uneven.

We propose an new LG model called the slow reaction (SR) model to simulate this effect. Our SR model is based on the spatial distribution of adjacent pedestrians. The sphere of action is within one cell distance, i.e. 0.4m. Leaving just one cell from the  front one, the pedestrian will get into slow reaction state. If the gap is more than one cell or zero, the target pedestrian will get into the free state or block state as the definition of 2.1 subsection. We introduce a parameter called the probability of slow reaction  $p_s$ to describe the quickness degree of reaction to front pedestrians. If someone gets into the slow reaction state, he will not follow the front moving pedestrian timely. The pedestrian could move forward with probability $p_s$ or remain at the same position with probability $1-p_s$ . Especially, the SR model becomes the standard LG model at $p_s=1$. The position update of SR model are calculated according to the following equations:
\begin{equation}
x_i^{n + 1} = \left\{ {\begin{array}{*{20}{c}}
{x_i^n,}&{d_i^n = 0,}&{{\rm{block}}\;{\rm{state}}}\\
{x_i^n + {p_s},}&{d_i^n = 1,}&{{\rm{slow}}\;{\rm{reaction}}\;{\rm{state}}}\\
{x_i^n + 1,}&{{\rm{other}},}&{{\rm{free}}\;{\rm{state}}}
\end{array}} \right.
\end{equation}
where $x_i^n$ and $v_i^n$ denote the position and velocity of pedestrian $i$ at time $n$ respectively. $d_i^n$ denotes the gap  between two successive pedestrians (not including the length of pedestrian) and $p_s$ is the probability of slow reaction within the limit of $0\leq p_s\leq1$.

\section{Simulations}
In Seyfried's experiment \cite{Seyfried2005_P10002}, the testers should have been distributed uniformly before starting to move.  However, due to the spatial discretization of LG model, it is difficult to distribute all virtual testers uniformly in the grid. It is impossible when the total number of pedestrians $N\geq22$ under the length of passageway $l_p=43$cell. We adopt a simple method of initialization at the beginning of simulation. We put all testers one by one with no gap from the left boundary of virtual passageway. We define the circle from the header entering the upper edge of measured section (17cell $\rightarrow$18cell) to the last pedestrian going out of the lower edge (22cell $\rightarrow$23cell). To reduce the effect of nonuniform distribution, we only use the data of velocity and density from the 50th circle to 100th circle to obtain the fundamental diagram.

We firstly investigate the relation between velocity and density. Through the computer simulation, we can obtain the global velocity and density in the whole passageway as well as the local velocity and density in the measured section. However, as no global velocity data was reported in \cite{Seyfried2005_P10002}, our global data cannot be compared. Therefore, we only compare the local velocity and density with \cite{Seyfried2005_P10002}.

Similar to the statistic method of \cite{Seyfried2005_P10002}, we calculate the individual velocity ${v_i} = {l_m}/(t_i^{out} - t_i^{in})$, where $t_i^{in}$ and $t_i^{out}$ is the entrance and exit times of measured section for pedestrian $i$. The mean velocity of one cycle $\bar v = \frac{1}{N}\mathop \sum \limits_{i = 1}^N {v_i}$ . The momentary density $\rho \left( t \right) = \sum\limits_{i = 1}^N {{\Theta _i}} \left( t \right)/{l_m}$ , where
\begin{equation}
{{\rm{\Theta }}_i}\left( t \right) = \left\{ {\begin{array}{*{20}{c}}
{\begin{array}{*{20}{c}}
{\frac{{t - t_i^{in}}}{{t_{i + 1}^{in} - i_i^{in}}}:t \in [t_i^{in},t_{i + 1}^{in}]}\\
{1:t \in [t_{i + 1}^{in},t_i^{out}]}
\end{array}}\\
{\begin{array}{*{20}{c}}
{\frac{{t_{i + 1}^{out} - t}}{{t_{i + 1}^{out} - t_i^{out}}}:t \in [t_i^{out},t_{i + 1}^{out}]}\\
{0:{\rm{otherwise}}}
\end{array}}
\end{array}} \right\}
\end{equation}

\begin{figure}[!ht]
\centering
\includegraphics[width=0.5\textwidth]{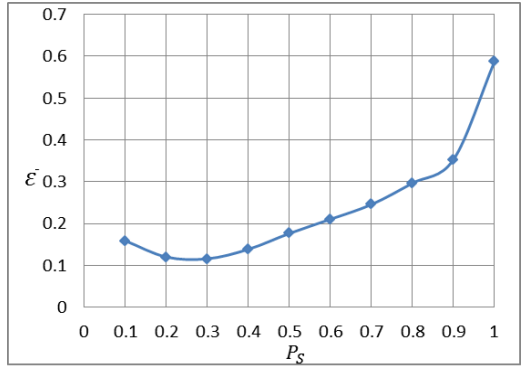}
\caption{The Root-Mean-Square error $\bar \varepsilon$ as the function of $p_s$.}
\label{fig7}
\end{figure}

The cycle density $\bar \rho $ is the mean value of the momentary density $\rho(t)$ during one cycle. Because \cite{Seyfried2005_P10002} only reported the results with six different total number of pedestrians $N=$ 1, 15, 20, 25, 30 and 34, we simulate five scenes ($N=$ 15, 20, 25, 30 and 34) to compare with the empirical data in \cite{Seyfried2005_P10002}. We set $p_s$ from 0.1 to 1 with increasing step $\Delta p_s = 0.1$ at each simulation implement. We calculate the value of density and velocity in each cycle from the 50th  to 100th circle and get the mean value. The Root-Mean-Square (RMS) error $\bar \varepsilon$ is calculated as follows:
\begin{equation}
\bar \varepsilon  = \sqrt {\frac{1}{5}\mathop \sum \limits_{k = 1}^5 {{({{\bar v}_k} - v_k^{man})}^2}}
\end{equation}
where $k$ represents the five scenes, ${\bar v_k}$ is the mean velocity of cycles from our simulation and $v_k^{man}$ is the mean velocity from the manual procedure in \cite{Seyfried2005_P10002}. The RMS error  $\bar \varepsilon$ as the function of $p_s$ is shown in Fig.~\ref{fig7}. We find the RMS error has the least value 0.12 at $p_s=0.3$ and the greatest value 0.58 at $p_s=1$. In addition, the RMS errors between $\bar{\rho}$ and $\rho^{man}$ from \cite{Seyfried2005_P10002} are nearly the same regardless of $p_s$ value.

\captionsetup[table]{labelsep=space}
\floatsetup{capposition=top}
\begin{table}[!ht]
\caption{The mean values and standard deviations ($\sigma$) of local density and individual velocity in measured section, which is gained from the video extraction by the manual procedure \cite{Seyfried2005_P10002}.}
\label{tab1}
\begin{tabular}{cccc}\hline
$N$	 & $\rho$[1/m] & 	$\rho^{man}$[1/m] & $v^{man}$[m/s] \\\hline
1		 &                       &                                    &1.24(0.15)           \\
15	 &0.87	           &0.77(0.12)	              &0.90(0.05)            \\
20	 &1.16	           &1.07(0.11)	              &0.56(0.05)             \\
25	 &1.45	           &1.39(0.12)	              &0.34(0.04)             \\
30	 &1.73	           &1.71(0.17)              &0.23(0.03)              \\
34	 &1.97	           &1.76(0.24)	              &0.17(0.03)              \\\hline
\end{tabular}
\end{table}

\floatsetup{capposition=top}
\begin{table}[!ht]
\caption{The mean values and standard deviations ($\sigma$) of local density and individual velocity in measured section by simulation when $p_s=1$ (standard LG model). The unit of simulation have been transformed to physical unit.}
\label{tab2}
\begin{tabular}{ccccc}\hline
$N$	 & $\rho$[1/m]  &  $\bar{\rho}$[1/m]  &  $\bar{v}$[m/s]  &  $\triangle v$[m/s] \\\hline
15	 &0.87	           &1.09(0.00)&1.24(0.00)&0.34\\
20	 &1.16	           &1.16(0.00)	&1.24(0.00)&0.68\\
25	 &1.45	           &1.47(0.05)	&1.04(0.05)&0.70\\
30	 &1.73	           &1.76(0.06)	&0.87(0.05)&0.64\\
34	 &1.97	           &1.99(0.02)&0.67(0.03)&0.50 \\\hline
\end{tabular}
\end{table}

\floatsetup{capposition=top}
\begin{table}[!ht]
\caption{The mean values and standard deviations ($\sigma$) of local density and individual velocity in measured section by simulation when $p_s=0.3$ (SR model). The unit of simulation have been transformed to physical unit.}
\label{tab3}
\begin{tabular}{ccccc}\hline
$N$	 & $\rho$[1/m]  &  $\bar{\rho}$[1/m]  &  $\bar{v}$[m/s]  &  $\triangle v$[m/s] \\\hline
15	 &0.87	           &0.87(0.02)&1.15(0.03)&	0.25 \\
20	 &1.16	           &1.19(0.05)	&0.61(0.03)&0.05 \\
25	 &1.45	           &1.44(0.02)	&0.36(0.01)&0.02 \\
30	 &1.73	           &1.74(0.03)&	0.20(0.01)&0.03 \\
34	 &1.97	           &1.98(0.04)&	0.12(0.01)&0.05  \\\hline
\end{tabular}
\end{table}

We compare our simulation results with the empirical data (see Table 1 from \cite{Seyfried2005_P10002}) and make a summary of two groups of results with parameters $p_s=1$ and $p_s=0.3$ in the Table 2 and Table 3 respectively. For Table 2, the residual error $\Delta \bar v$ is the greatest at $N=25$ and the smallest at $N=15$ among five scenes. For Table 3, the residual error $\Delta \bar v$ is the smallest at $N=25$ and the greatest at $N=15$. At low density ($N=15$), the difference between the mean velocity of SR model and standard LG model is not significant. But the mean gap ($\bar{d}=1/\bar{\rho}$) is a little longer ($1cell\leq \bar{d}\leq2cell$) in SR model than in LG model. At high density ($N=$30 and 34), the block states dominate the pedestrian flow for the two models. At medium density ($N=$20 and 25), the moving states of pedestrians in SR and LG model have important difference. For example, all pedestrians is in free states in LG model at $N=20$. However, a part of pedestrians is in slow reaction states or block states in SR model. As an extreme case for $p_s=0$, no pedestrian can move forward in SR model if $N\geq22$. Only leaving two cells in the front can a pedestrian move in free state in SR model.

\floatsetup{capposition=bottom}
\begin{figure}[!ht]
\centering
\subfloat[]{\includegraphics[width=0.45\textwidth]{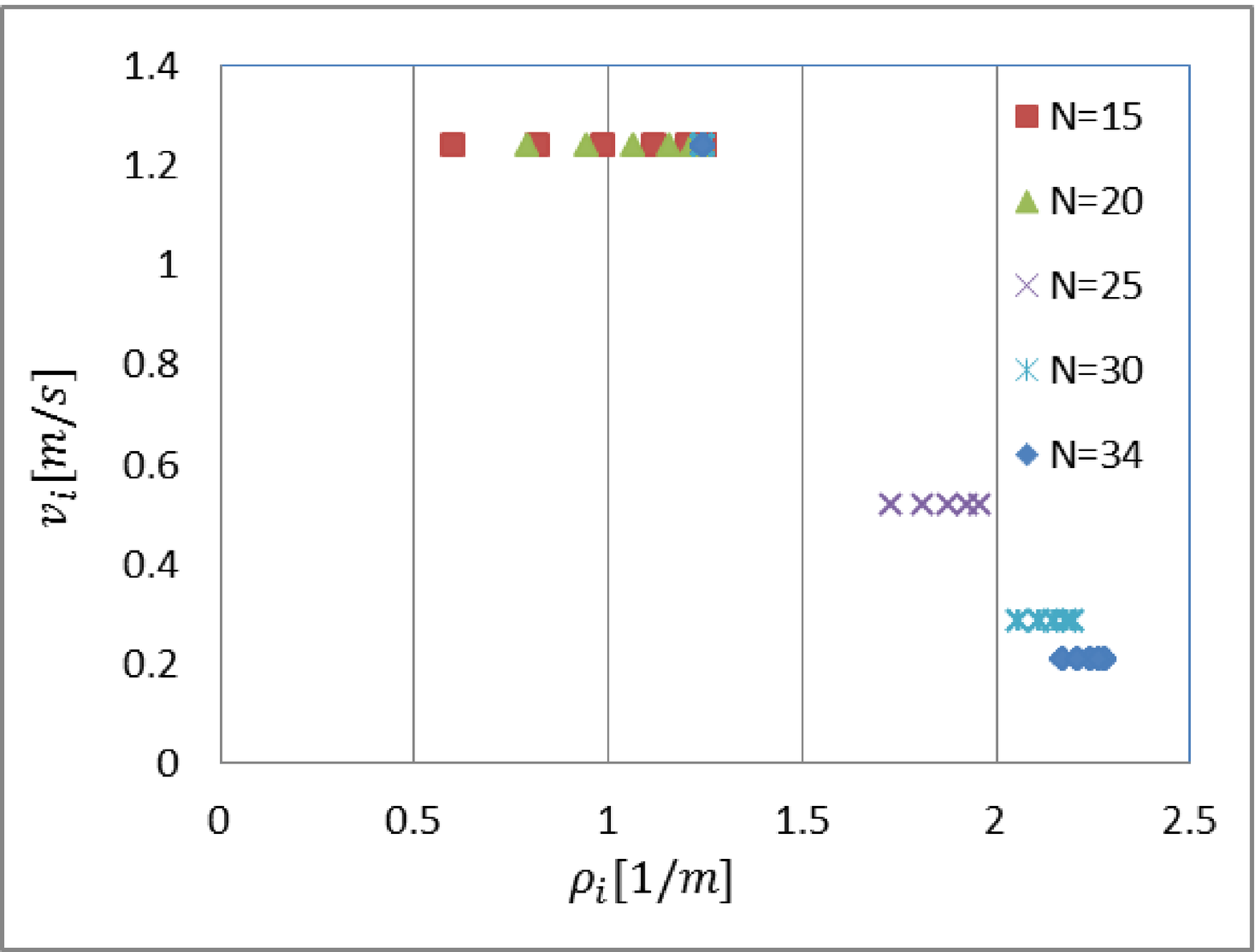}}\hspace{10pt}
\subfloat[]{\includegraphics[width=0.45\textwidth]{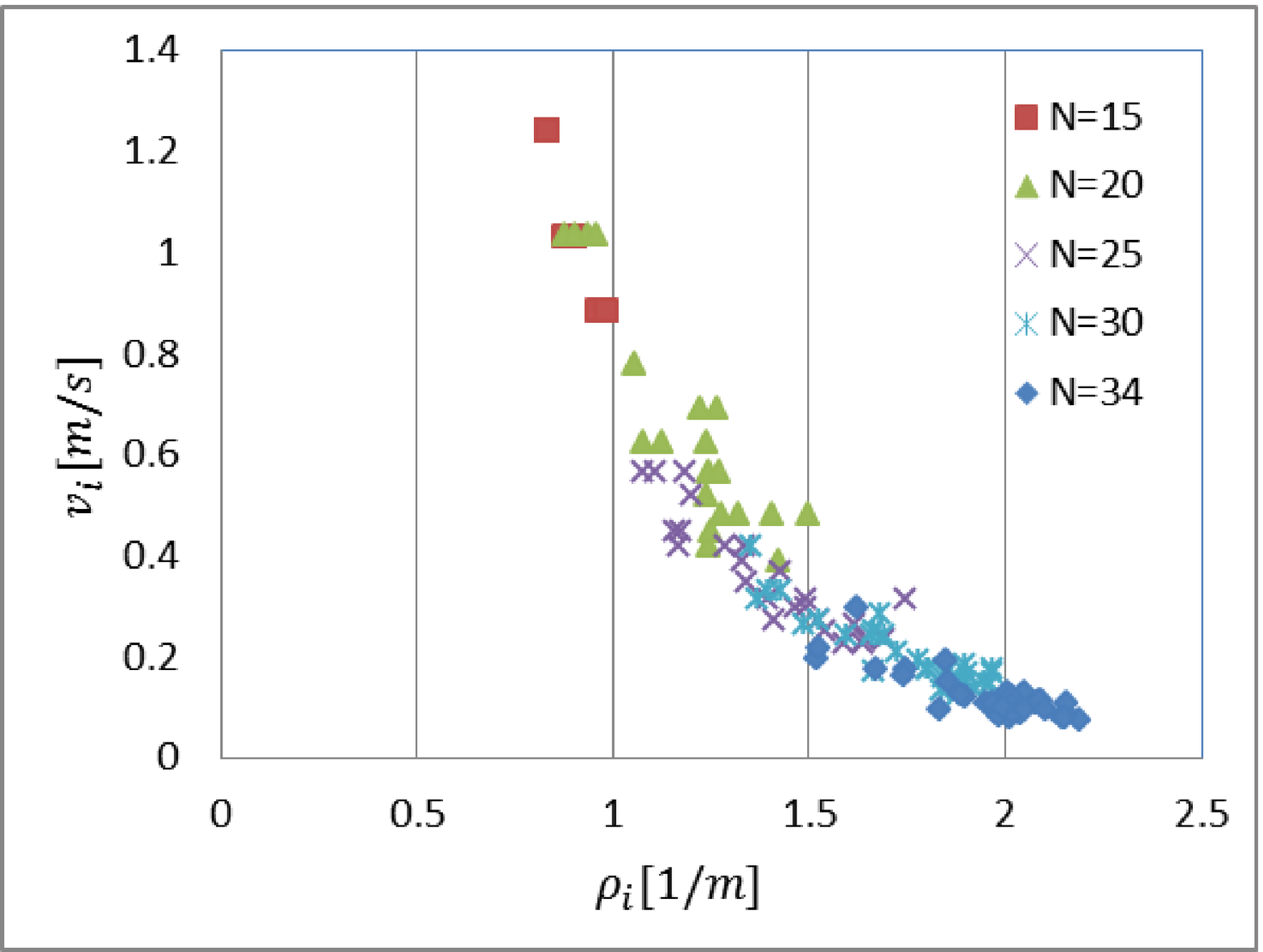}}\\
\subfloat[]{\includegraphics[width=0.45\textwidth]{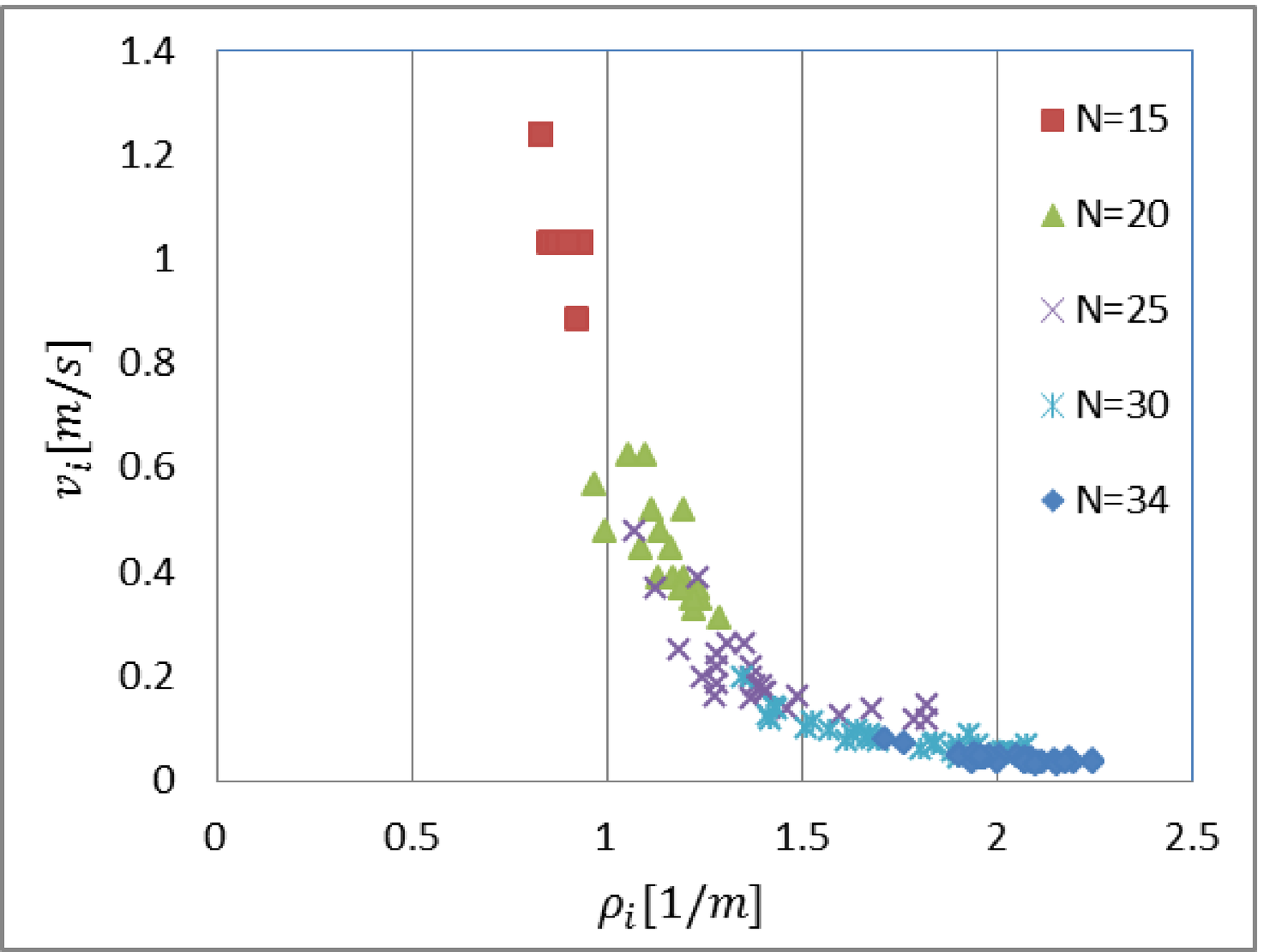}}\hspace{10pt}
\subfloat[]{\includegraphics[width=0.45\textwidth]{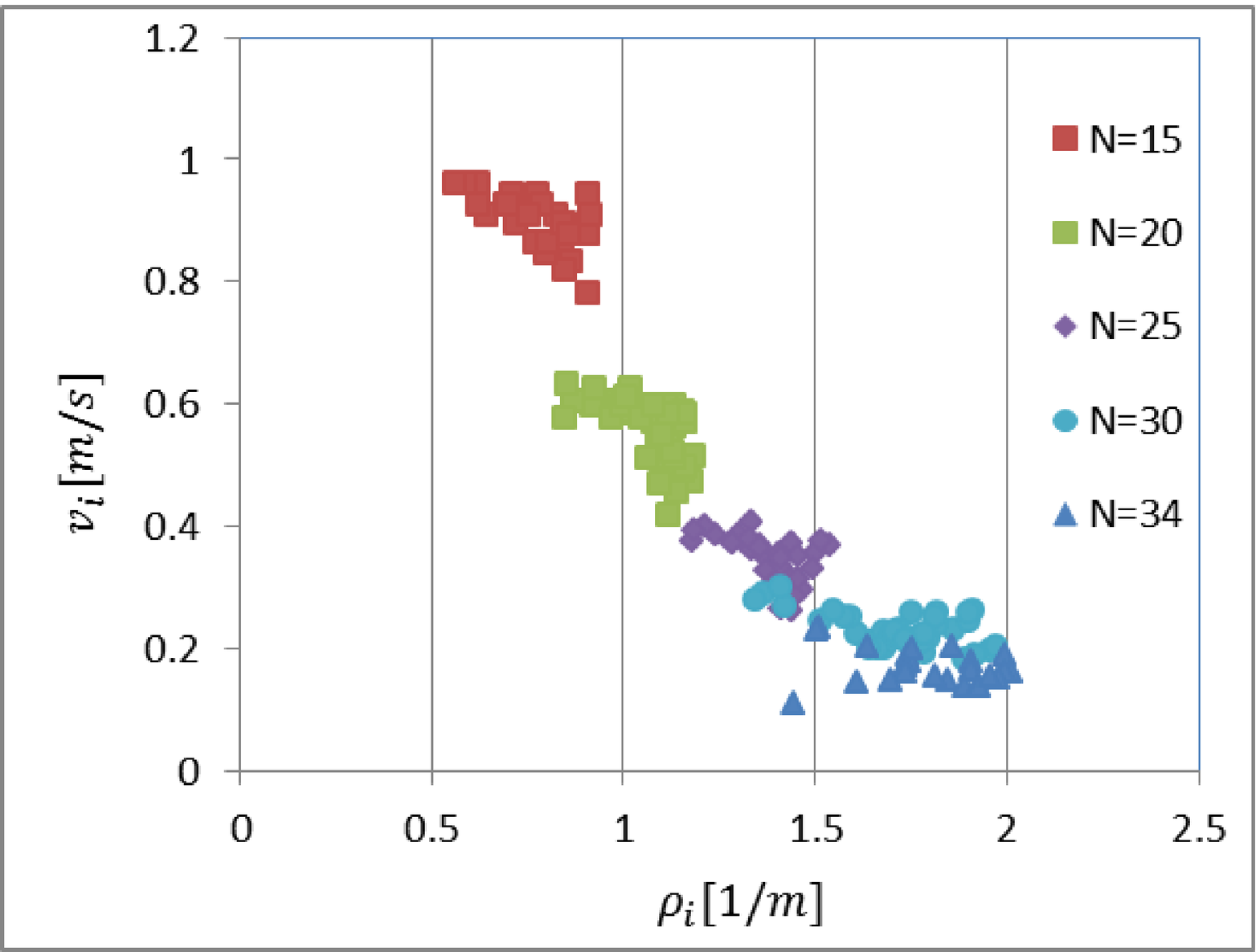}}
\caption{Dependency between the individual velocity and density of the 100th cycle with $N=$ 15, 20, 25, 30 and 34. (a) $p_s=1$ (standard LG model); (b)$p_s=0.3$ (SR model); (c) $p_s=0.1$ (SR model); (d) Seyfried's experimental data in \cite{Seyfried2005_P10002}.}
\label{fig8}
\end{figure}

To validate the distribution of individual velocity and density at fundamental diagram, we draw velocity-density diagrams with values of $p_s=$1, 0.3 and 0.1. The local density $\rho_i$ of pedestrian $i$  is the mean value of the  momentary density $\rho(t)$ during the time-slice $[t_i^{in},t_i^{out}]$. The ($v_i,\rho_i$) distribution of standard LG model in Fig.~\ref{fig8}(a) is very different from Fig.~\ref{fig8}(b), Fig.~\ref{fig8}(c) and Fig.~\ref{fig8}(d). The LG model with $p_s=1$ become a deterministic discrete model and everyone's individual velocity at steady state is fixed without fluctuation. For each density, the individual velocities distribute along a line. All of the individual velocities of $N=$15 and 20 are same, which reach the free speed (1.24m/s). This is because the critical density $\rho_c=1.27$[1/m] ($N=22$) in our simulation system, after which the stop-and-go waves occur and phase separation is observable. The discussion about critical density and stop-and-go waves are set up later.
In \cite{Seyfried2005_P10002}, the density-velocity points for the cycles with $N=$ 15, 20 and 25 distribute in clearly separated areas and these areas blend for the cycles with $N=$ 30 and 34 (see Fig.~\ref{fig8}(d)). However, the values of individual velocity distribute broader in our simulation than in \cite{Seyfried2005_P10002}. There are no obvious boundaries between the density-velocity points of two different $N$ in Fig.~\ref{fig8}(b) and Fig.~\ref{fig8}(c). A possible explanation is that the slow reaction gap with only one cell between the front and back pedestrian is not long enough. One observes that the values of individual velocity at low density $N=$ 15 and 20 in simulation are higher than that of empirical data. The phenomenon is due to the original discreteness of LG model in terms of velocity and position. In our model, a pedestrian either moves one cell with $v_i (t)=1$cell (1.24m/s in physical unit) or stays still with $v_i (t)=0$cell. There is on mechanism to generate an intermediate value 0cell$<v_i (t)<$1cell about the momentary individual velocity.

\begin{figure}[!ht]
\ffigbox[\textwidth]{}
{
\begin{subfloatrow}
\ffigbox[\FBwidth]{\caption{}}{\includegraphics[width=0.5\textwidth]{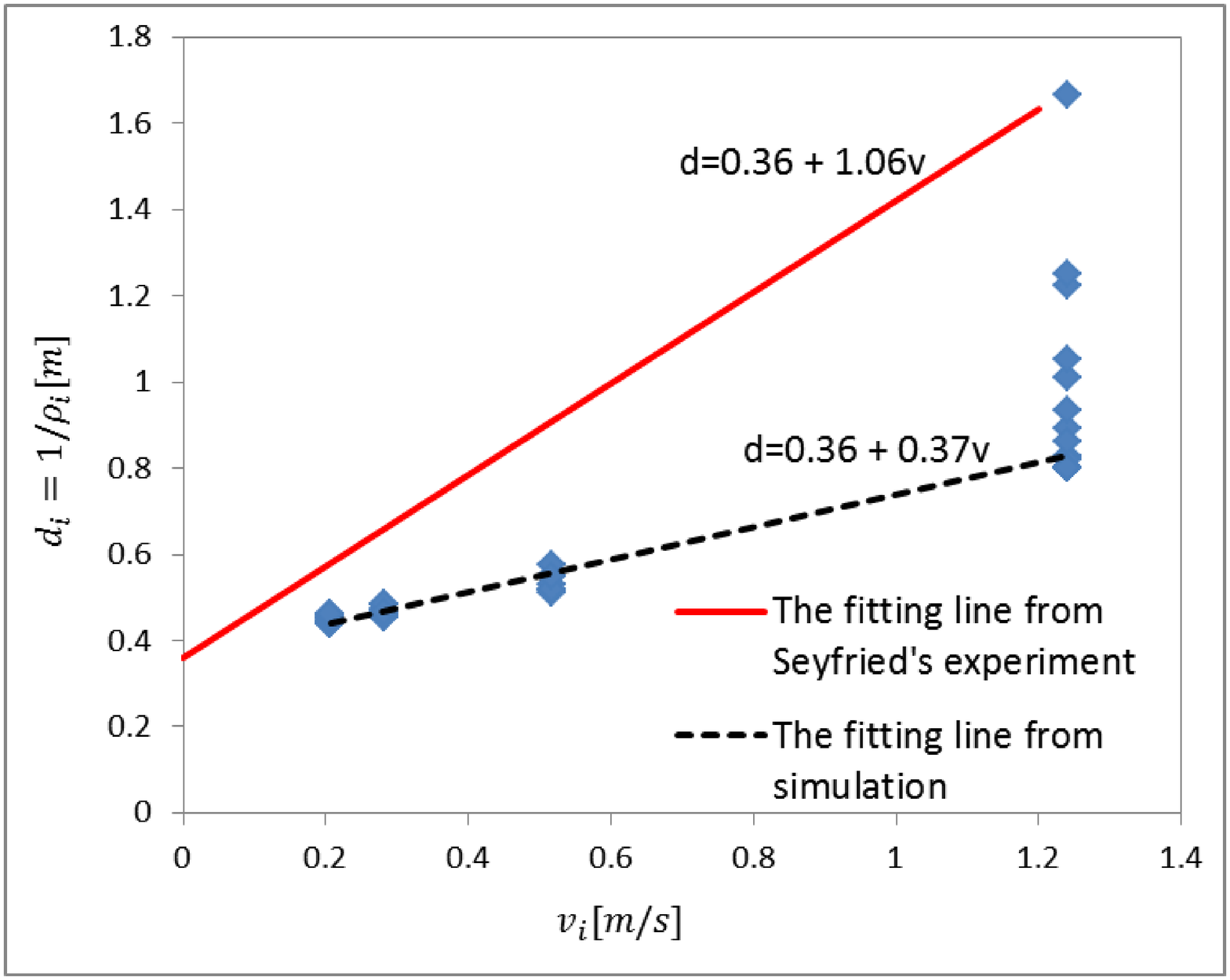}}
\hspace{10pt}
\ffigbox[\FBwidth]{\caption{}}{\includegraphics[width=0.5\textwidth]{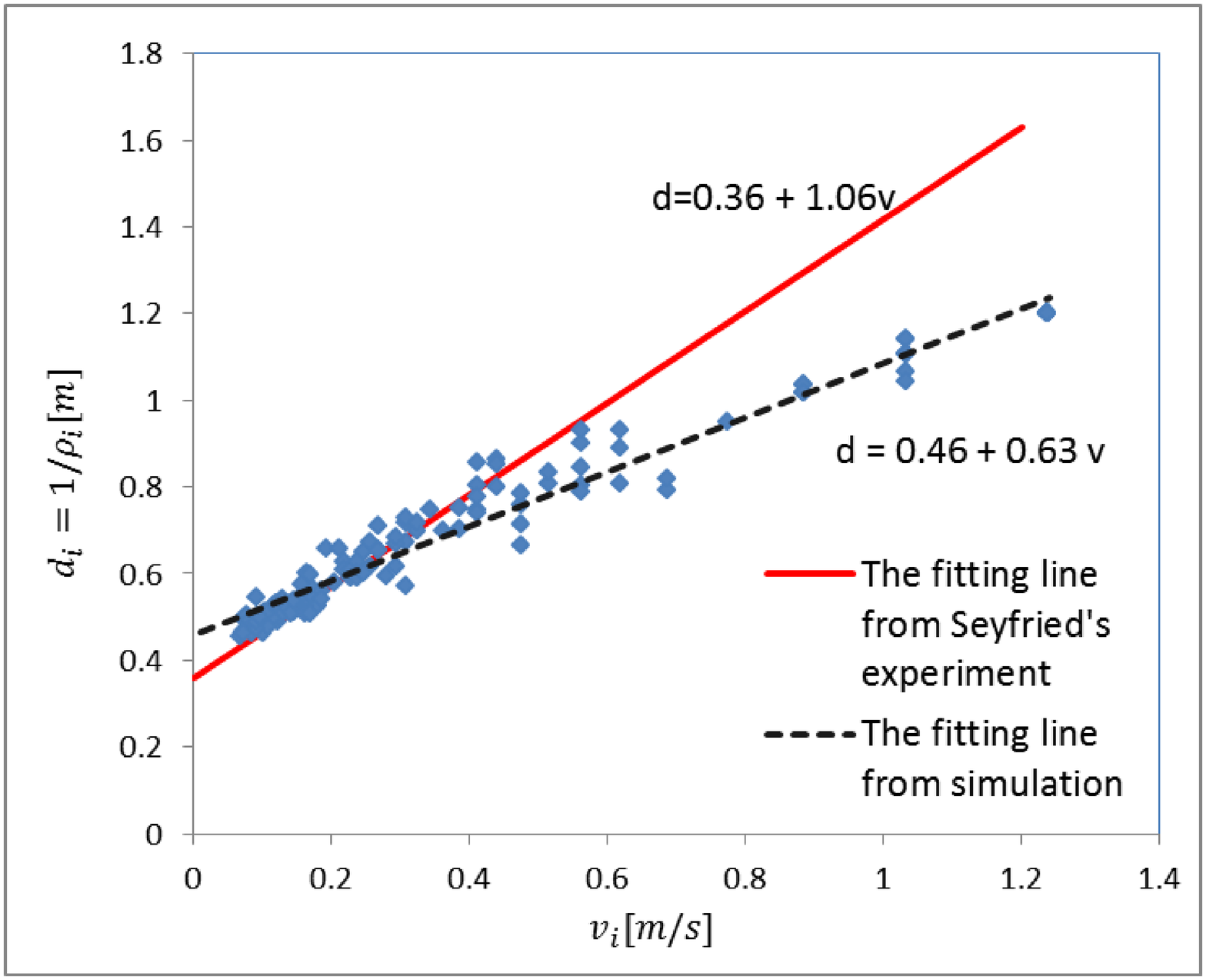}}
\end{subfloatrow}
\caption{Dependency between required length and velocity of the 100th cycle with $N=$ 15, 20, 25, 30 and 34. (a) $p_s=1$; (b) $p_s=0.3$.}
\label{fig11}
}
\end{figure}

In Fig.~\ref{fig11}(a), the individual velocities do not distribute widely in $d_i - v_i$ diagram but arrange in several lines. In Fig.~\ref{fig11}(b), a linear relationship with $d=0.46+0.63v$ gives the best fit to our simulation data with coefficient of determination $R^2=0.96$. Although the linear trend agrees with \cite{Seyfried2005_P10002}, the slope 0.63 of fitting line is a little lower than the slope 1.06 in Seyfried's experiment. One possible explanation is that the size of cell is square with 0.4m*0.4m. In real life, the shape of pedestrian is an approximate ellipse with a typical value of 0.58m*0.33m \cite{Still2001}. Another reason is that due to spatial discretization of LG model, pedestrians cannot be distributed uniformly in the grid.

\begin{figure}[!ht]
\centering
\subfloat[]{\includegraphics[width=0.5\textwidth]{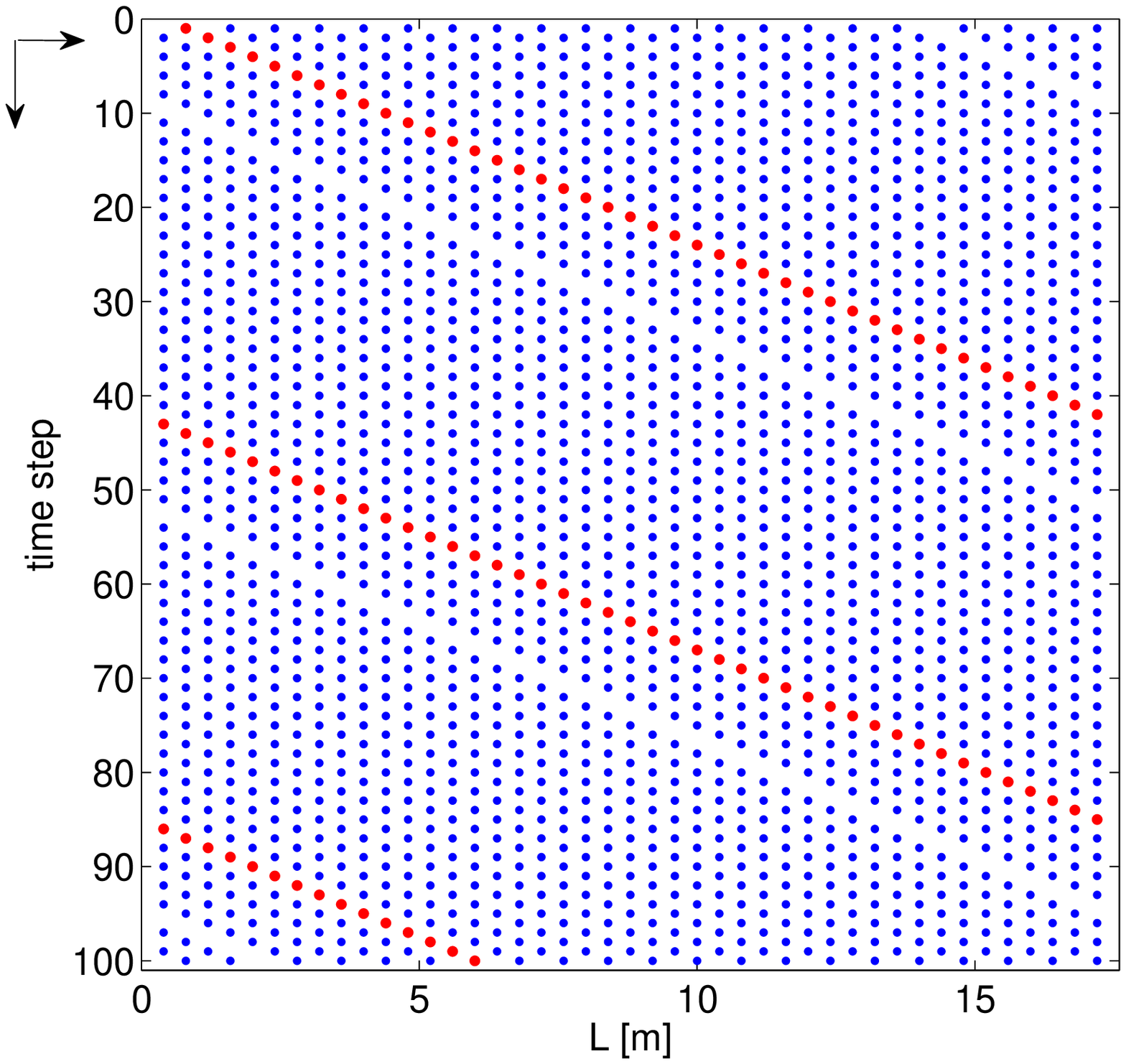}}
\subfloat[]{\includegraphics[width=0.5\textwidth]{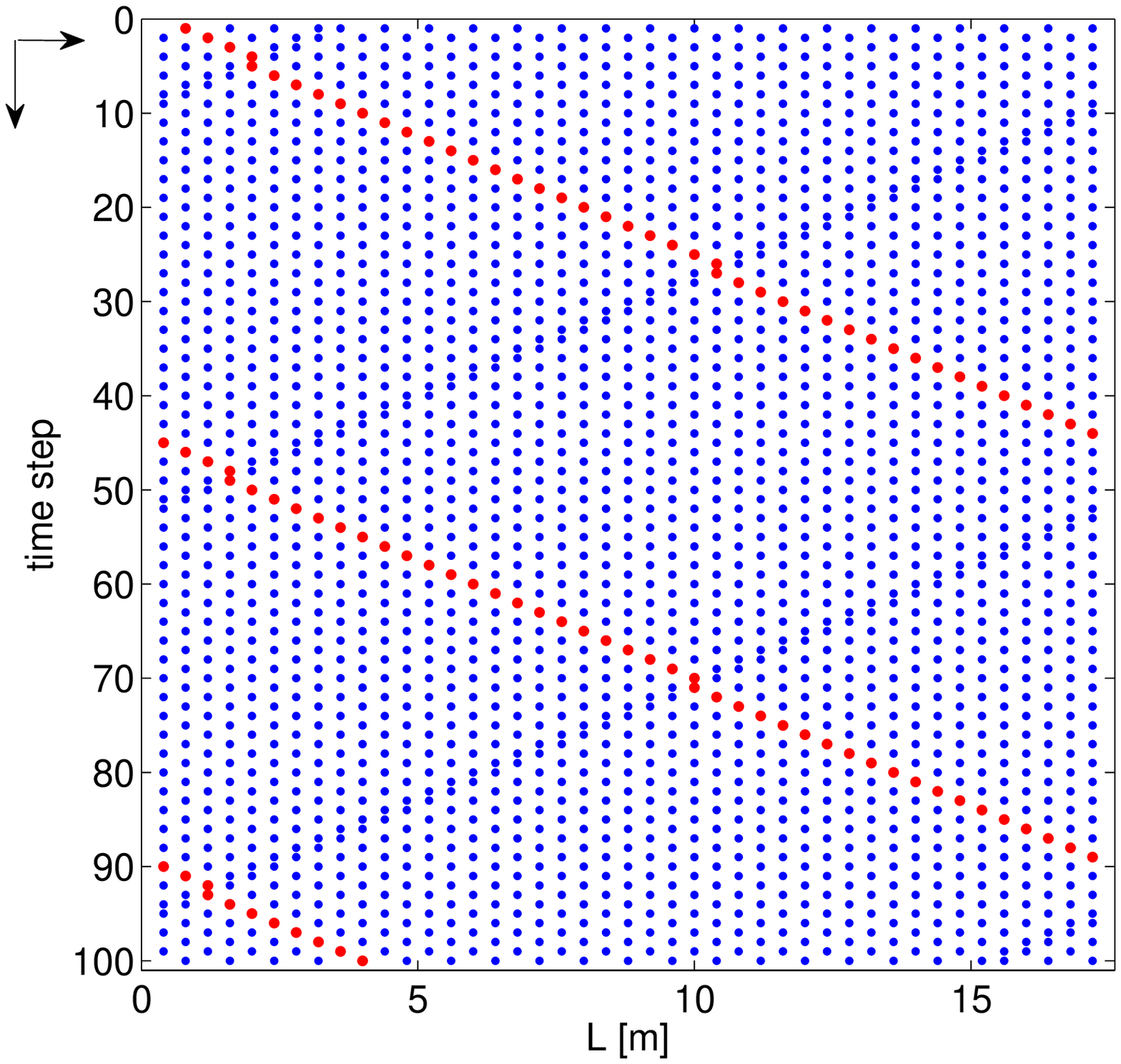}}\\
\subfloat[]{\includegraphics[width=0.5\textwidth]{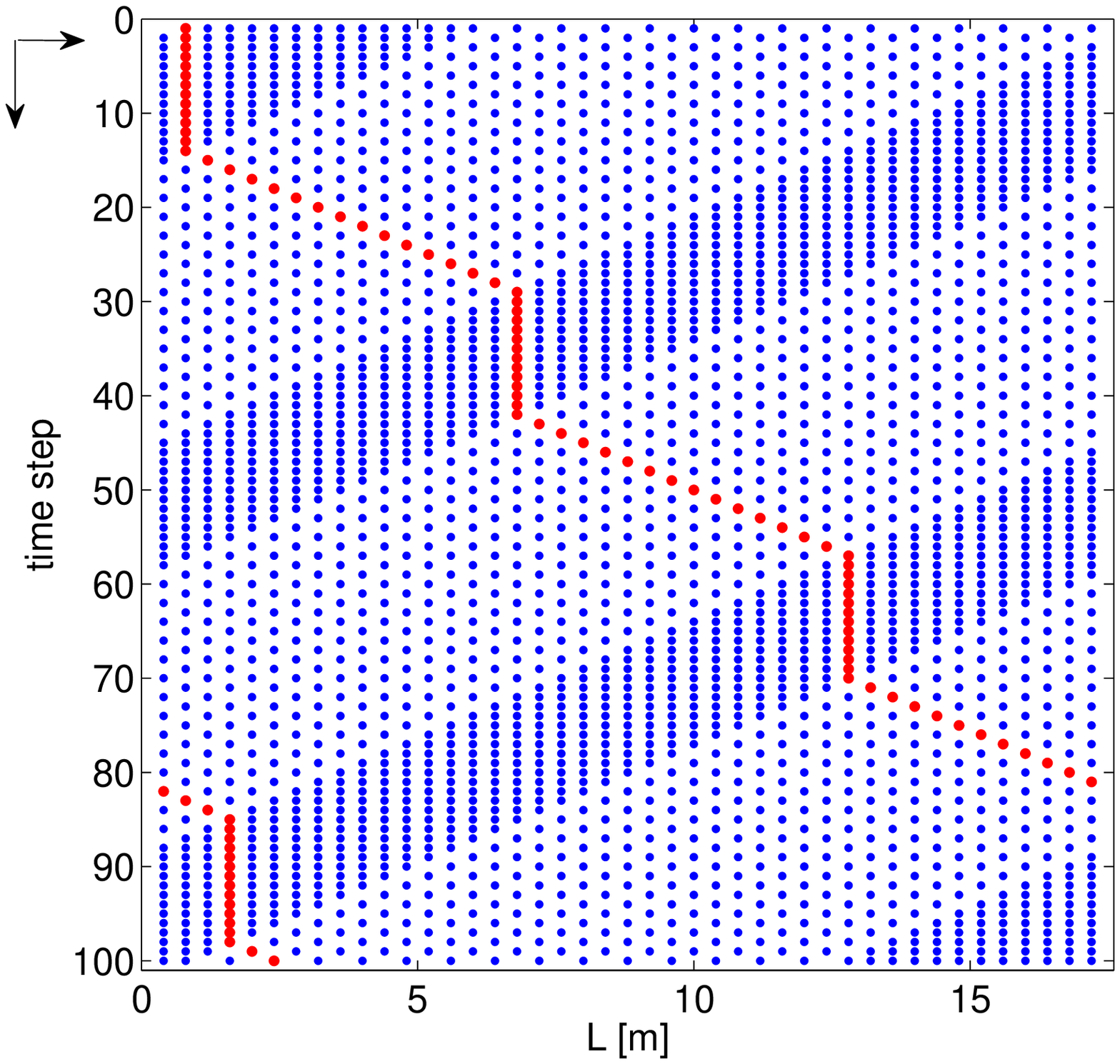}}
\subfloat[]{\includegraphics[width=0.5\textwidth]{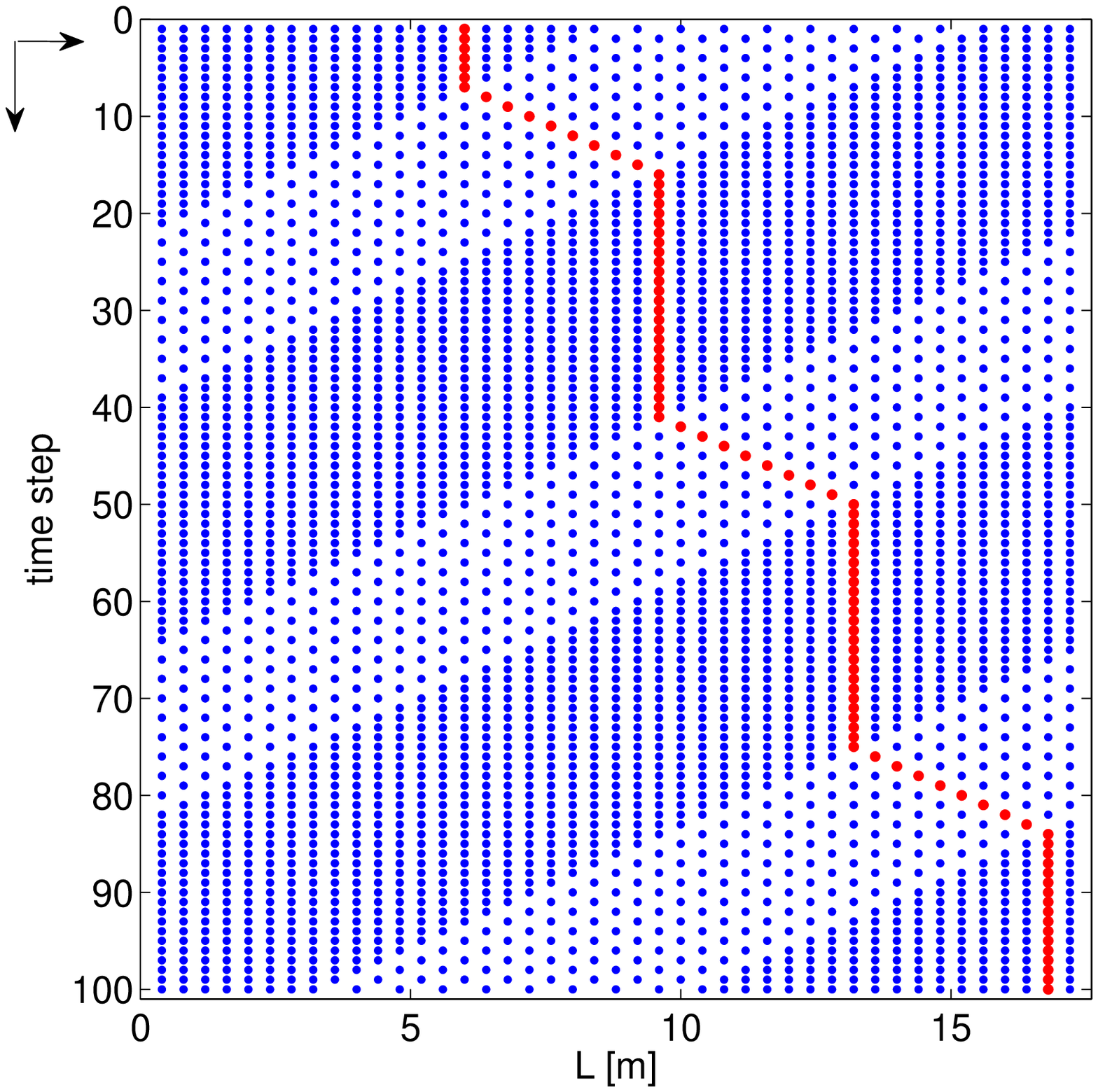}}
\caption{(Colored) The propagation (rightward) of stop-and-go waves in the pedestrian flow from time step 5,000$\sim$5,100 when $p_s=1$ (standard LG model). The series of red dots denotes the trajectory of a pedestrian during this time. (a) $N=21$; (b) $N=22$; (c) $N=28$; (d) $N=34$. }
\label{wave_1}
\end{figure}

The stop-and-go waves and phase separation in pedestrian dynamic of standard LG model is shown in Fig.~\ref{wave_1}. With the compression effect of stop-and-go wave, the pedestrian flow is divided into two phase: the free phase and jam phase. The common border of the two phases goes upstream circularly with propagation velocity $v_p=-1$cell/step (i.e. -1.24m/s in physical unit and minus denotes the opposite direction from pedestrian flow). The stop-and-go waves in LG model is similar to the stop-and-go waves of highway traffic, which often happens at the upper stream of highway bottleneck \cite{Kerner2008_215101}. Although the stop-and-go wave and phase separation in the single-file pedestrian flow was not reported in \cite{Seyfried2005_P10002}, the same phenomena were observed and discussed in the field experiments of \cite{Seyfried2010_496}. A series of larger scale field experiments were arranged by the same research team in the wardroom of Bergische Kaserne D\"{u}sseldorf at 2006 \cite{Seyfried2010_496}. The group of test persons was composed of soldiers instead of students. The experimental set-up is similar to that of \cite{Seyfried2005_P10002} except the size of passageway and the number of testers. In addition, similar result was reported in simulation by a social force model in \cite{Seyfried2005_357}. We observed in simulation that the critical density $\rho_c=1.27$[1/m] (corresponding $N=22$ in $L=1.73$m), after which the stop-and-go waves are observable. This coincides with the simulation result of $\rho_c=1.21$[1/m] in \cite{Seyfried2005_357}. However, one can not observe in standard LG model that some pedestrians leave much larger than average gaps in front, which was reported in \cite{Seyfried2005_357}.

\begin{figure}[!ht]
\centering
\subfloat[]{\includegraphics[width=0.5\textwidth]{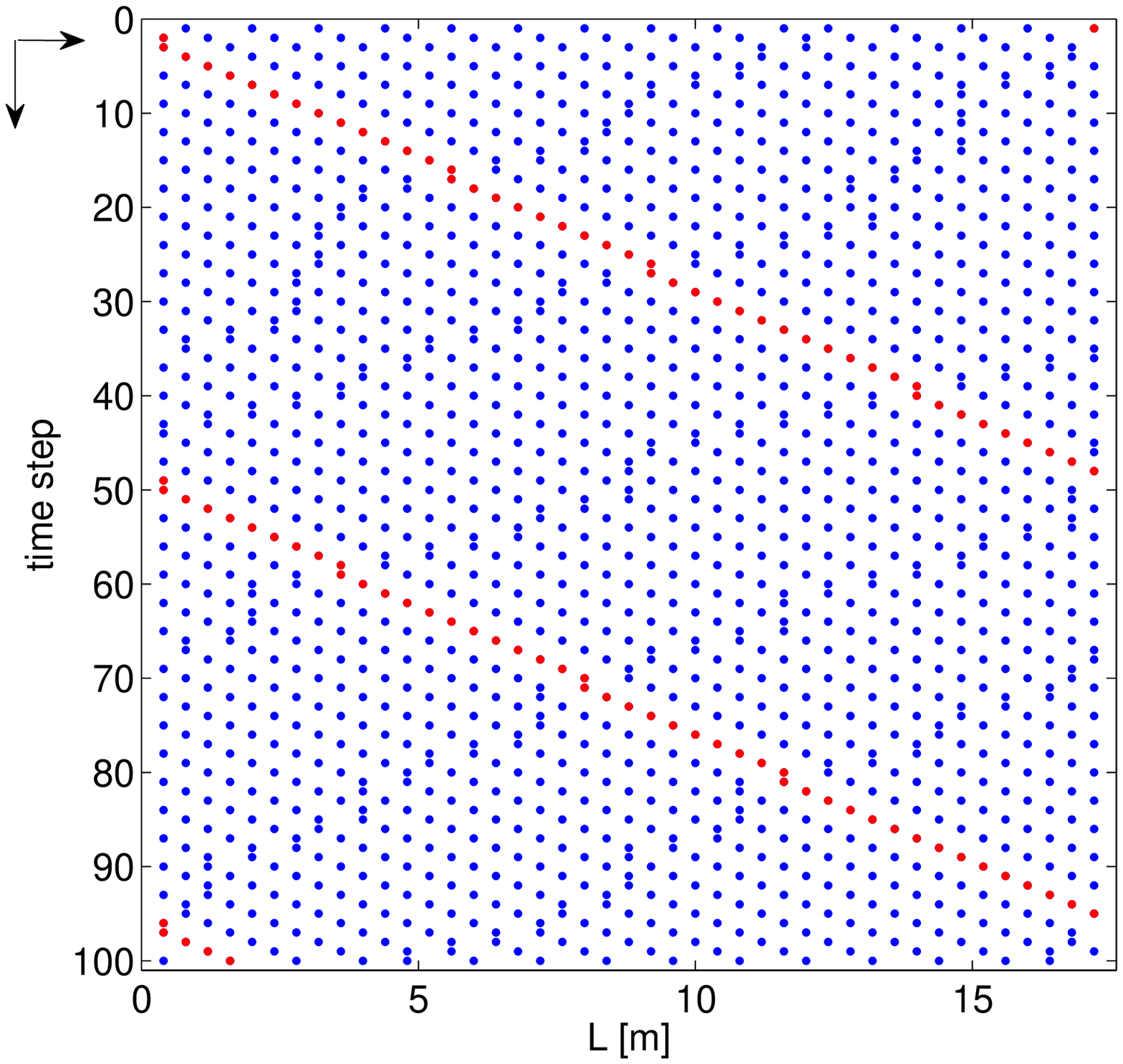}}
\subfloat[]{\includegraphics[width=0.5\textwidth]{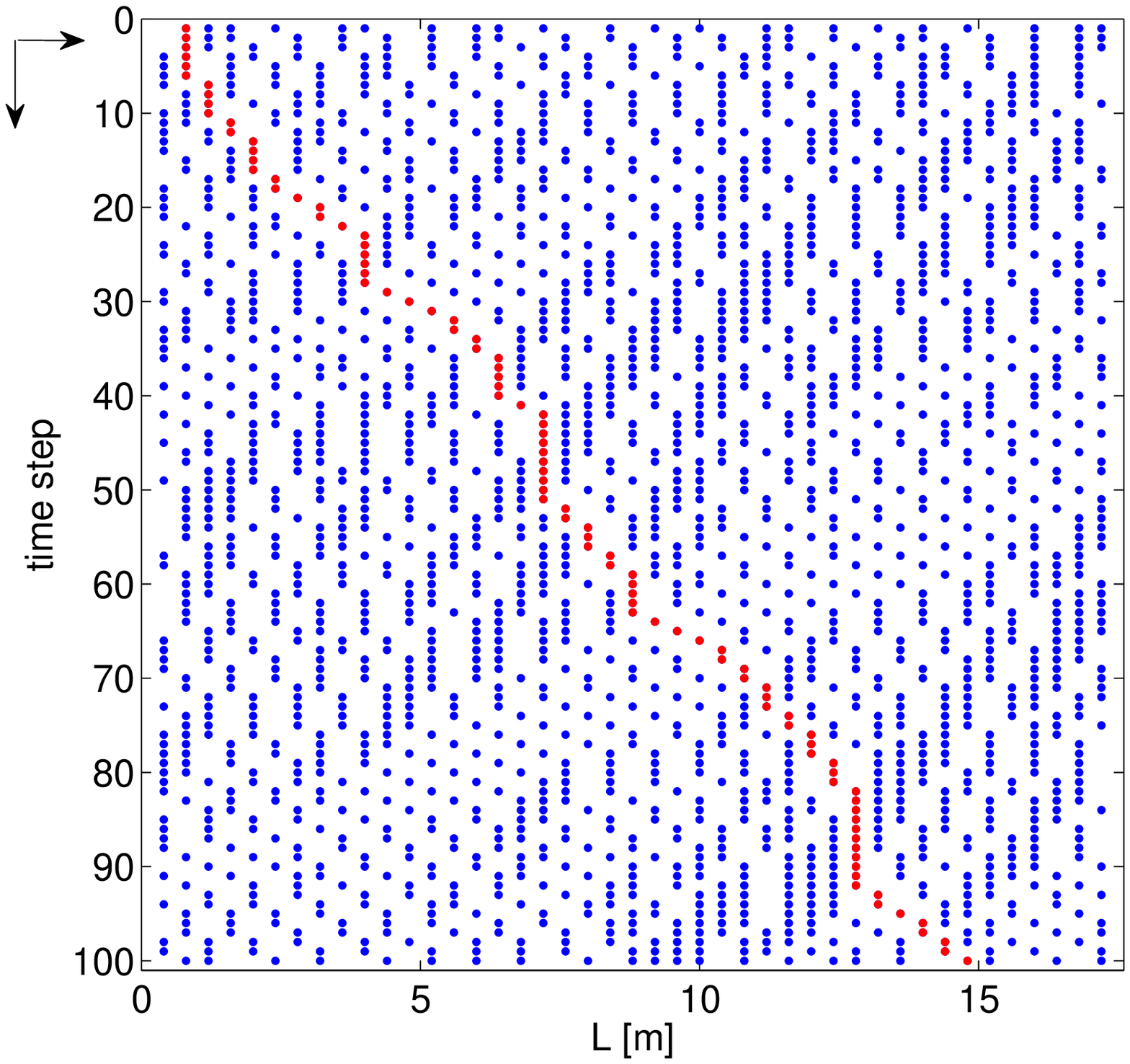}}\\
\subfloat[]{\includegraphics[width=0.5\textwidth]{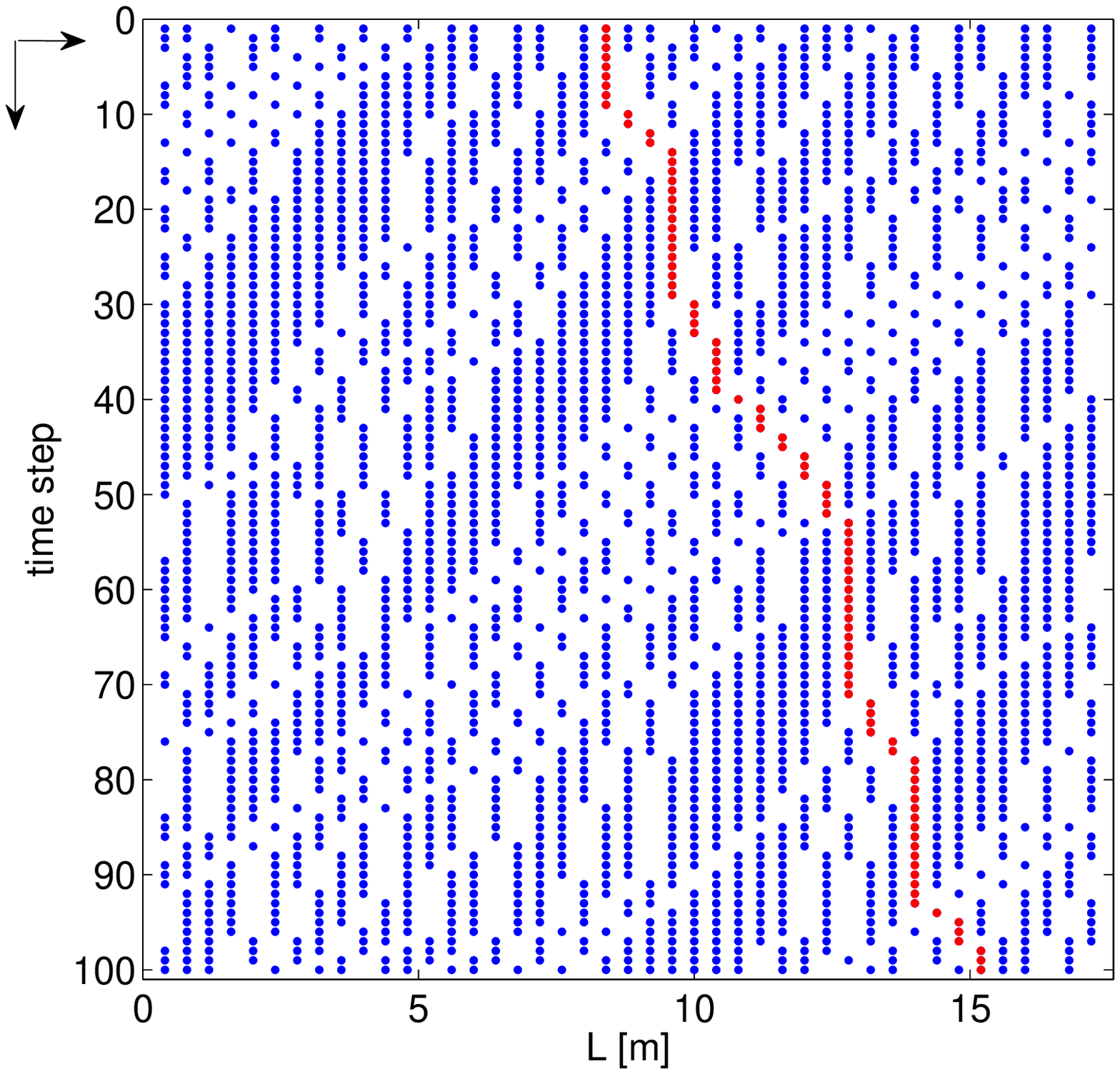}}
\subfloat[]{\includegraphics[width=0.5\textwidth]{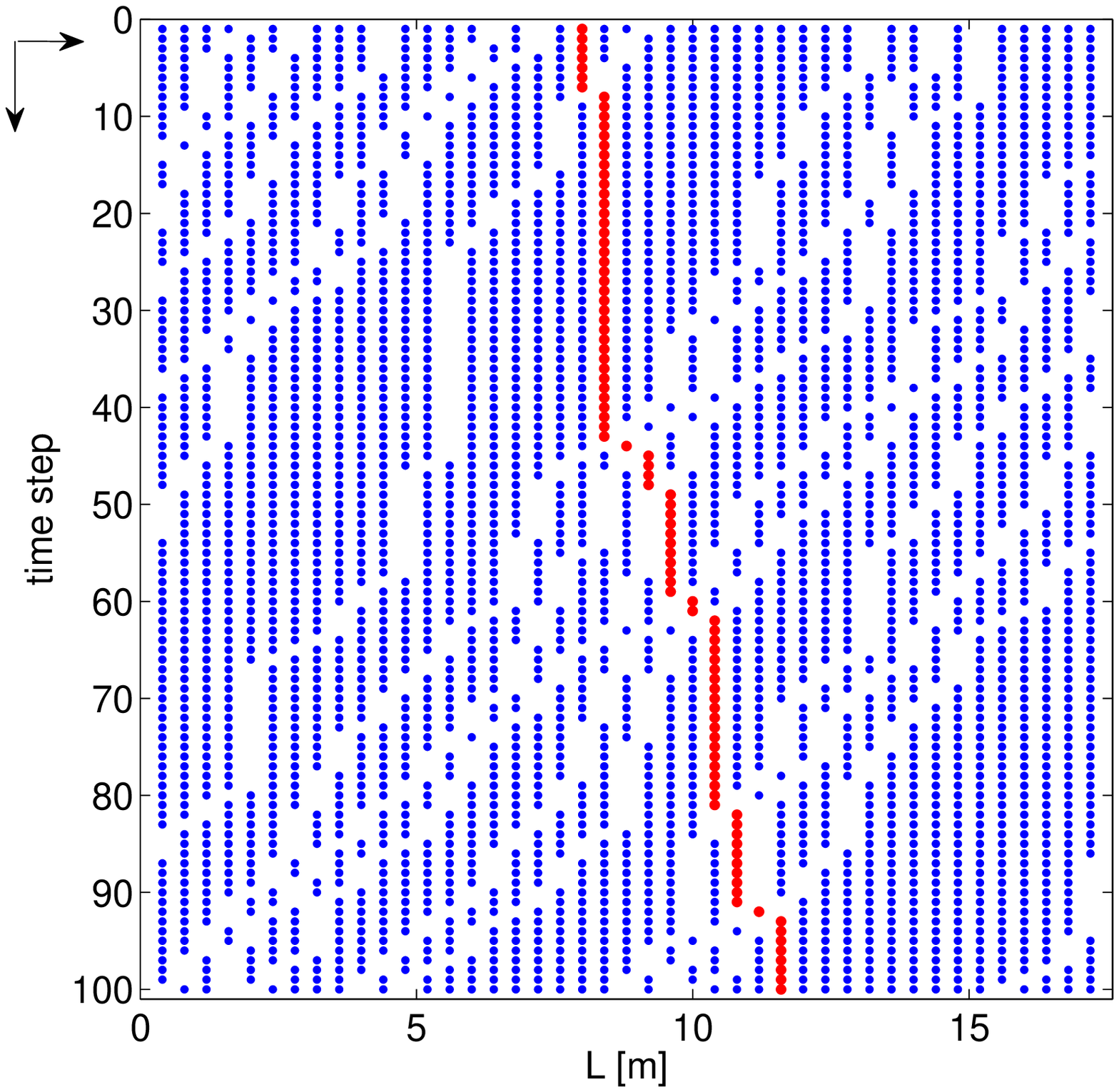}}
\caption{(Colored) The propagation (rightward) of stop-and-go waves in the pedestrian flow from time step 5,000$\sim$5,100 when $p_s=0.3$ (SR model). The series of red dots denotes the trajectory of a pedestrian during this time. (a) $N=15$; (b) $N=22$; (c) $N=28$; (d) $N=34$. }
\label{wave_2}
\end{figure}
The stop-and-go waves and phase separation for SR model is shown in Fig.~\ref{wave_2}. The global stop-and-go wave observed in standard LG model does not exist in SR model. Instead, several small density waves go upstream with propagation velocities $|v_p|<1$ . For $N>14$, the stop-and-go waves occur and phase separation is observable. The pedestrian moving in free state must leave two empty cells at least in front of itself.  Therefore, the critical density $\rho_c=43/3=14.3$cell in theory. One can observe in SR model that some pedestrians leave much larger than average gaps in front, which coincides with \cite{Seyfried2005_357}. The pedestrian flow is jostled by the stop-to-go waves and forms several jam clusters separated by one or two empty cells. The fluctuation of local density and individual velocity indicates the inner randomness of pedestrian movement and is in accordance with the observations in \cite{Seyfried2005_P10002} and \cite{Seyfried2010_496}.

\section{Conclusions}
In this work, we propose a model called the slow reaction model (SR model) to describe the pedestrian's delayed reaction in single-file movement. We simulate and reproduce Seyfried's field experiment  at the research centre J\"{u}lich in 2005 \cite{Seyfried2005_P10002} and use its empirical data to validate our SR model. We compare the simulation result of SR model with that of the standard LG model. Our simulation work includes: (a) the fundamental diagram about the relation between one-dimensional density and velocity; (b) dependency between the individual required length and velocity; (c) stop-and-go waves and phase separation. We test different probability of slow reaction $p_s$ in SR model and found the simulation data of $p_s=0.3$ fit the empirical data \cite{Seyfried2005_P10002} best. The RMS error of mean velocity between simulation and empirical data is the smallest ($\bar \varepsilon=0.12$) at $p_s=0.3$. The RMS error of SR model $p_s=0.3$ is smaller than that of standard LG model $p_s=1$ ($\bar \varepsilon=0.58$).
 In the range of $p_s=0.1\sim0.3$, our fundamental diagram between  velocity and density by simulation coincides with field experiments in \cite{Seyfried2005_P10002}. The distribution of  individual velocity in fundamental diagram in SR model agrees with the empirical data \cite{Seyfried2005_P10002} better than that of standard LG model. In addition, we observe the stop-and-go waves and phase separation in pedestrian flow by simulation, which were observed in \cite{Seyfried2005_357} and \cite{Seyfried2010_496}. We reproduced the phenomena of uneven distribution of interspaces by SR model while the standard LG model did not implement. The uneven distribution of interspaces was also reproduced by social force model in \cite{Seyfried2005_357}. The fluctuation of local density and individual velocity indicates the inner randomness of pedestrian movement. The SR model can reproduce the evolution of spatio-temporal structures of pedestrian flow with higher fidelity to Seyfried's experiments \cite{Seyfried2005_P10002} than the standard LG model.

There are some imperfections for current SR model. The spatial partition is not fine enough and the expression of  velocity lacks the middle levels. It is hard to describe the individual momentary velocity between 0m/s to 1.24m/s. This mechanism will make the mean individual velocity larger at low density and smaller at high density than that of the empirical data. Although we introduce the slow reaction mechanism and calculate the mean velocity of several cycles to alleviate this problem, this imperfection has not been solved fully until now. Introducing a finer discretization of the space for formulating pedestrians' behavior in SR model is is worth further research.

\section*{Acknowledgments}
We are grateful to the ped-net-group (www.ped-net.org) for providing useful empirical data. We use the data that come from the project funded by the German Science Foundation (DFG) under DFG-Grant No. KL 1873/1-1 and SE. 1789/1-1.

\bibliographystyle{model1-num-names}
\bibliography{Reference}

\end{document}